\documentclass{PoS}

\title{Decay constants from twisted mass QCD.}

\ShortTitle{Decay constants from twisted mass QCD.}

\author{P. Dimopoulos\\
         Dip. di Fisica, Universit{\`a} di Roma Tor Vergata and INFN,
      Sez. di Tor Vergata,\\ Via della Ricerca Scientifica, I-00133 Roma, Italy \\
      E-mail:  \email{dimopoulos@roma2.infn.it}}

\author{\speaker{C. McNeile}%
         \thanks{On behalf of the ETM Collaboration}\\
Department of Physics and Astronomy\\
The Kelvin Building\\
University of Glasgow, Glasgow G12 8QQ, U.K.\\
        E-mail: \email{c.mcneile@physics.gla.ac.uk}}

\author{C. Michael\\
        Theoretical Physics Division, Dept of Mathematical Sciences,
 University of Liverpool, Liverpool L69 3BX, U.K.\\
        E-mail: \email{cmi@liverpool.ac.uk}}

\author{S. Simula\\
        INFN, Sez. di Roma Tre,
        Via della Vasca Navale 84, I-00146 Roma, Italy \\
        E-mail: \email{simula@roma3.infn.it}}

\author{C. Urbach\\
Humboldt-Universit\"{a}t zu Berlin,  Institut f\"{u}r Physik
Mathematisch-Naturwissenschaftliche Fakult\"{a}t I\\
Theorie der Elementarteilchen / Ph\"{a}nomenologie\\
Newtonstr. 15, 12489 Berlin Germany
        E-mail: \email{Carsten.Urbach@physik.hu-berlin.de}}

\abstract{
We present results for chiral extrapolations of the mass and decay
constants of the rho meson. The data sets used are the $n_f$=2 unquenched gauge
configurations generated with twisted mass fermions by the European Twisted
Mass Collaboration. 
We describe a calculation of three
decay constants in charmonium and explain why they are required.
}

\FullConference{The XXVI International Symposium on Lattice Field Theory\\
		 July 14-19 2008\\
		 Williamsburg, Virginia, USA}

\begin{document}

\section{Introduction}

The twisted mass formalism~\cite{Frezzotti:2000nk,Frezzotti:2003ni} 
has proven itself to be a powerful
tool to extract precision physics in the 
pseudo-scalar~\cite{Boucaud:2007uk,Boucaud:2008xu}
and baryon sector~\cite{Alexandrou:2008tn}. 
In this work we look at one of the simplest resonances the $\rho$
meson. The additional complication for the $\rho$ meson is that it
decays via the strong interaction to two pions. In this project we aim
to study the effect of the decay of the $\rho$ mass and decay constant
via the chiral extrapolation formulae.  As an extension of the study of the
decay constants of the light vector meson we also studied the decay
constants of the $J/\psi$ and $\eta_c$ mesons.

The twisted Wilson action with the tree level Symanzik
gauge action was used (see~\cite{Boucaud:2008xu} for details).
In this analysis we use the $n_f$ = 2 unquenched data sets at 
$\beta$=3.9 $24^3 \; 48$ and 
$\beta=4.05$ $32^3 \; 64$ with lattice spacings 
determined from $f_\pi$ to be
0.0855 fm and 0.0667 fm respectively~\cite{Dimopoulos:2007qy}. 
These lattice
spacings were consistent with those from the 
mass of the nucleon~\cite{Alexandrou:2008tn}.

\section{Mass of the $\rho$ meson}

The masses and decay constants for the light vector meson were
obtained by fitting an order 4 smearing matrix to a factorising fit
model~\cite{Boucaud:2008xu}. The correlators were generated with the 
"one-end" trick to improve the signal to noise ratio. We only consider
unitary data. Vladikas et al.~\cite{LAT08VLADIKAS}
studies an independent set of vector meson
correlators with partial quenching on the same set of gauge
configurations.

In this section we discuss the chiral extrapolation of the 
mass of the $\rho$ meson.
Bruns and Mei{\ss}ner~\cite{Bruns:2004tj}
have published a "new'' chiral extrapolation formulae for
the $\rho$ meson, based on
a modified $\overline{MS}$ regulator.
\begin{equation}
M_\rho = M^0_\rho + c_1 M_\pi^2  + c_2 M_\pi^3 + 
c_3 M_\pi^4 \ln ( \frac{M_\pi^2}{M_\rho^2} )
\label{eq:BrunsMass}
\end{equation}
It is important 
to check that size of $c_i$ from the fits to the lattice 
data is consistent 
with other estimates from phenomenology.
Bruns and Mei{\ss}ner~\cite{Bruns:2004tj}
claim that phenomenology prefers $\mid c_i \mid < 3 $. 
Leinweber
et al.~\cite{Leinweber:2001ac} 
claim to know sign of $c_2$ (negative) and magnitude
from  $\omega \rho \pi$ coupling.
The original effective field theory calculation
Jenkins, et al.~\cite{Jenkins:1995vb} had $c_3$=0.
There are too many parameters to determine from our data, so
we use an augmented $\chi^2$ with the above constraints
built in~\cite{Lepage:2001ym} 
$$
\chi^2_{aug} = \chi^2 + \sum_{j=2}^3 \frac{ (c_i -0)^2 } {3^2} 
$$

We also tried the fit model suggested by
Leinweber et al.~\cite{Leinweber:2001ac}.
The formulae doesn't obey the power counting of effective field theory, 
but does include an explicit term for the decay of the 
$\rho$  to two mesons. Also some coefficients are fixed
from experiment in the fit model.

Although we have data at two lattice spacings we only fit
the $\beta$ = 3.9 $24^3\; 48 $ data sets. Unfortunately the 
$\rho$ correlators at $\beta=4.05$ and the $32^3 \; 64$ $\beta=3.9$
data were too noisy to be useful.
See~\cite{McNeile:2007fu} for a comparison between our results
for the mass of the $\rho$ meson and other lattice collaborations.
 We are investigating various
measurements techniques, such as colour diluted stochastic sources,
to reduce the statistical  errors.

\begin{figure}
\centering
\includegraphics[%
  scale=0.5,
  angle=270,
  origin=c,clip]{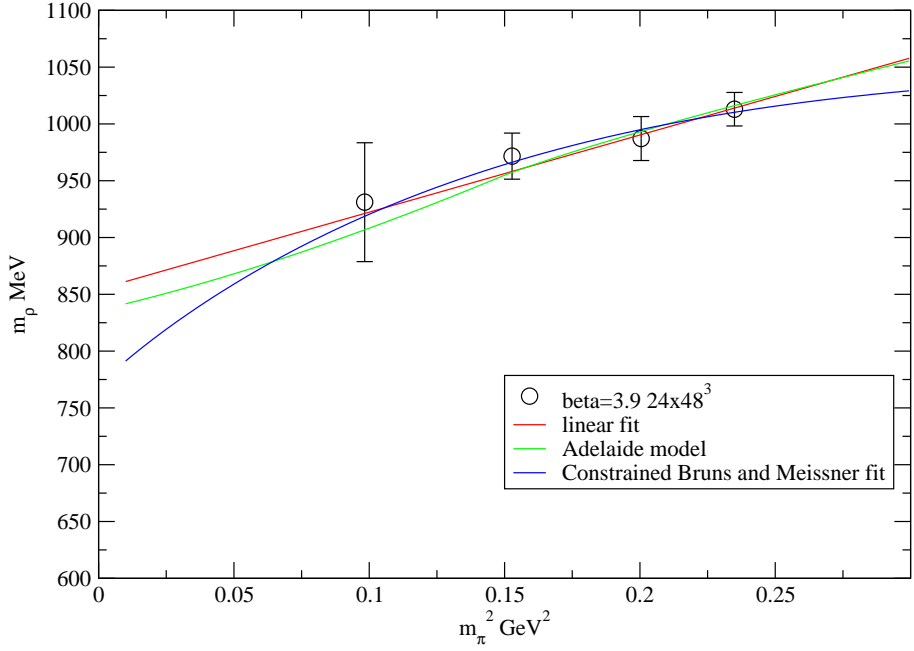}
\vspace{-30pt}
\caption{Chiral extrapolation of the mass of the $\rho$ meson}
\label{fig:mvPLOT}
\end{figure}

The lattice data are shown in figure~\ref{fig:mvPLOT}.
A fit linear in the square of the pion mass gives
$m_\rho$ = 867(29) MeV ($\beta=3.9$ data), compared
to the experimental value of 770 MeV.
Naively the $\rho$ data prefers a different lattice spacing to 
that from $f_\pi$, but we think that the error is due to missing 
chiral corrections.
We fit the Bruns-Mei{\ss}ner model to the data with constraints
on the $c_2$ and $c_3$ coefficients 
and obtain
the preliminary result is
$m_\rho$ = 807(8) MeV. For the 
Adelaide fit~\cite{Leinweber:2001ac}
method for the 
$\rho$ meson, we obtain $m_\rho$ = 847(72) MeV. 
The statistical errors
need to be reduced for a detailed
comparison between the lattice data and 
the results of effective field theory.

\section{Decay constants of the $\rho$ meson} \label{se:lightdecay}

Light cone sum rules require the transverse decay
constant of the $\rho$
meson~\cite{Ball:2006nr,Ball:2006eu}
for the extraction of $\frac{\mid V_{td} \mid}{\mid V_{ts} \mid } $
from the $B \rightarrow \rho \gamma $ and
$B \rightarrow K^\star \gamma $
decays, and other semileptonic decays of the 
B meson~\cite{Becirevic:2003pn}.

In the continuum the decay constant of the $\rho$ meson
is defined
via
\begin{equation}
\langle 0 \mid \overline{\psi}(x) \gamma_\mu \psi (x) \mid \rho \rangle = 
m_\rho f_\rho
\epsilon_\mu
\label{eq:rhodecayDEFN}
\end{equation}
The transverse decay constant ($f_V^T(\mu)$)
of the $\rho$ meson is defined
by
\begin{equation}
\langle 0 \mid 
\overline{\psi} \sigma_{\mu \nu} \psi
\mid \rho \rangle = 
i f_V^T(\mu) ( p_\mu \epsilon_\nu - p_\nu \epsilon_\mu )
\label{eq:TRANSrhodecayDEFN}
\end{equation}
where $\sigma_{\mu \nu} = i/2 [\gamma_\mu , \gamma_\nu] $.
The $f_V^T(\mu)$ decay constant can not be determined from
experiment.
For the charmonium part of this analysis we studied the 
decay constant of the pseudoscalar meson.
\begin{equation}
f_{PS} = ( 2 \mu) \frac{ \mid
\langle 0 \mid P^1(0) \mid P \rangle \mid }
{M_{PS}^2 }
\end{equation}

To renormalise the currents we use the 
results from the
Rome-Southampton method, slightly updated from
those reported by Dimopoulos et
al.~\cite{Dimopoulos:2007fn} last year.
We remind the reader that in twisted mass QCD the charged
vector current renormalises with $Z_A$.
The
renormalisation factor of the tensor current depends on the scale, so
we run to 2 GeV using the method described by Becirevic et
al.~\cite{Becirevic:2003pn}. We will include the results 
from Gracey's three loop calculation at a later time~\cite{Gracey:2003yr}.

\begin{table}[tb]
\centering
\begin{tabular}{|c|c|c|} \hline
$\beta$  &  $Z_A$  & $Z_T(\mu=1/a)$  \\  \hline
3.90     &  0.771(4)   & 0.769(4)  \\ 
4.05     &  0.785(6)   & 0.787(7) \\ \hline
\end{tabular}
\caption{Non-perturbative renormalisation factors used in this analysis}
\label{tb:Zfact}
\end{table}

The chiral perturbation theory calculations for
the chiral extrapolations of the leptonic vector decay 
constants have been calculated~\cite{Bijnens:1998di}.
The corrections
due loops start at $m_q \log m_q$ and 
$m_q^{3/2}$~\cite{Bijnens:1998di}.
Unfortunately our light decay constant data
is too noisy to look for chiral corrections, so we
use simple linear fits in the quark mass. The chiral
perturbation theory for tensor sources has been developed,
but no loop calculations are available~\cite{Cata:2007ns}.
Although by taking ratios of correlators the ratio 
of $\frac{f_\rho^T}{f_\rho}$  can be extracted 
directly~\cite{Becirevic:2003pn}, we prefer to 
separately compute  $f_\rho^T$  and $f_\rho$,
because $f_\rho$ is known from experiment (207 MeV) is
thus a good validation test that the $\rho$ to 2$\pi$ decay is 
under control. In table~\ref{tb:decaySUMM} we present our 
preliminary results for the light vector decay constants,
and compare to previous results in the literature.

\begin{table}[tb]
\begin{center}
\begin{tabular}{|c|c|c|c|c|} \hline
Group & Method & $f_\rho^T$(2 GeV) MeV & $f_\rho$ MeV&  $\frac{f_\rho^T}{f_\rho}$  \\ \hline
Becirevic et al.~\cite{Becirevic:2003pn}  & quenched  & 
150(5) & - & $0.72(2)^{+2}_{0}$ \\
Braun et al.~\cite{Braun:2003jg}  & quenched  & 
154(5) & - &  0.74(1) \\
QCDSF 2005~\cite{Gockeler:2005mh}  & unquenched  & 
168(3) & 256(9) & 0.66(3) \\
RBC-UKQCD~\cite{Allton:2008pn} & unquenched  & 143(6)  & - &  0.69(3) \\
\hline
This work (preliminary) & unquenched  & 177(26)  & 
229(35) &  0.76(14)
\\ \hline
\end{tabular}
\end{center}
\caption{Summary of results for transverse and leptonic decay constants
of the $\rho$ meson}
\label{tb:decaySUMM}
\end{table}

\section{Decay constants in charmonium}

One way to understand charmonium production and decay is to use the
NRQCD formalism, where non-perturbative information is encoded in a
few matrix elements. NRQCD mostly produces a good description of
experimental data. See~\cite{Brambilla:2004wf} for a review and a
discussion of the convergence of the velocity expansion in the charm
region.
However  
the results from  Belle~\cite{Abe:2002rb}
for double charmonium production
\begin{equation}
\sigma[ e^+ e^- \rightarrow J/\psi + \eta_c]
{\cal B}
= 25.6 \pm 2.8 \pm 3.4 \; fb
\end{equation}
where ${\cal B} < 1$,
are much larger than the prediction from leading order NRQCD by 
Bodwin et al.~\cite{Bodwin:2002kk}. 
\begin{equation}
\sigma[ e^+ e^- \rightarrow J/\psi + \eta_c]_{(NRQCD ; LO)}
= 3.78 \pm 1.26  \; fb
\end{equation}
BaBar has a similar result to Belle's for this 
process~\cite{Aubert:2005tj}.
Although further NRQCD calculations that include relativistic corrections
and higher order perturbative corrections have increased
the leading order NRQCD results, the final result
is still lower than the experimental 
result~\cite{Zhang:2005cha}.

It is claimed that calculations that use light cone wave 
functions of charm mesons agree with the Belle 
result~\cite{Bondar:2004sv}.
However, Bodwin~\cite{Bodwin:2005ec}
remarks that it is not
clear that the  model light cone wave
functions used are  close to true quarkonium wave functions.
Lattice QCD calculations should be able to constrain some of the 
parameters of the light cone wave function of the charmonium 
mesons. In particular the decay constants are parameters
of the light cone wave functions.
We refer the reader to figure 6  
in~\cite{Choi:2007ze} to see the effect of the decay
constants on the production cross-section. 
See~\cite{Chung:2008sk} for a 
recent review of developments in this field.

As stressed by the FNAL and 
HPQCD collaborations~\cite{Follana:2007zz},
the computation of the decay constant of the
$J/\psi$ meson, that is known from experiment, is an important validation test
for calculations that compute
decay constants that contain heavy quarks.

The lattice calculation is essentially the same as for
the light quarks in section~\ref{se:lightdecay}. 
As normal the charmonium correlators
are much less noisy than for the equivalent light quark correlators.
We used charged pseudoscalar interpolating operators, and
we do not include any disconnected diagrams in the analysis.
We computed the leptonic and transverse decay constant of the 
$J/\psi$ meson and the decay constant of the $\eta_c$ meson
for the $\beta$ values 3.9 and 4.05. We used three heavy quark masses
that interpolated the mass of the charm quark. We tuned the charm
mass by looking at the $J/\psi$ meson. Only local-local correlators
were used.

\begin{figure}
\centering
\includegraphics[scale=0.5,angle=270]{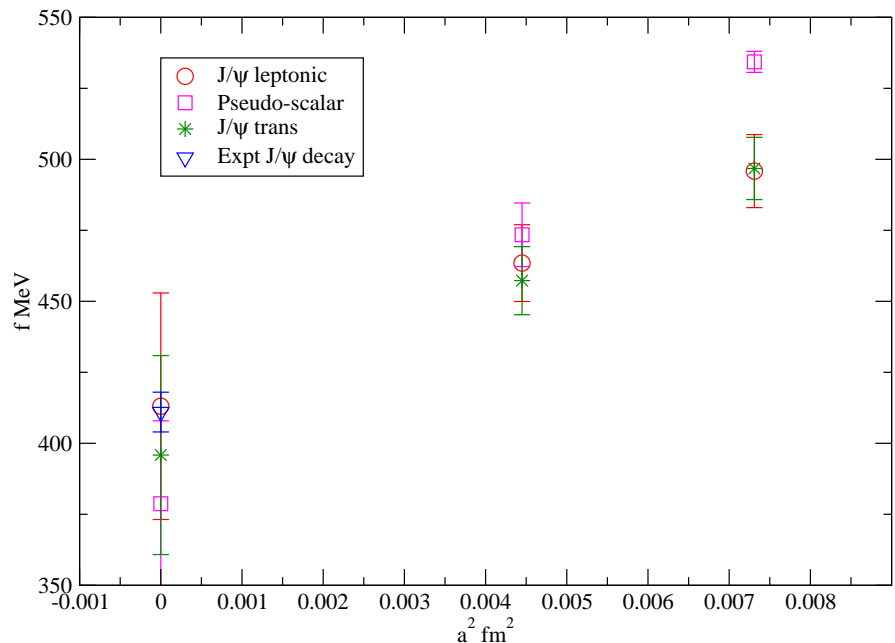}
\vspace{-30pt}
\caption{Decay constants as a function of square of lattice spacing}
\label{fig:decaycharm}
\end{figure}

The results for the three decay constants are plotted in 
figure~\ref{fig:decaycharm}.
We quote preliminary results:
$f_{J/\psi}$ = 413(40) MeV, 
$f_{J/\psi ; trans}$  =396(35) MeV,
and
$f_{\eta_c}$ =379(29) MeV,
for the decay constants in the continuum limit. The errors on the
final results are inflated because we only had data at two lattice
spacings.

The experimental value of $f_{J/\psi}$= 411(7) MeV
(see~\cite{Dudek:2006ej} for example).  There is a ``sort of
experiment result'' $f_{\eta_c}$ = 335(75) MeV from $B \rightarrow
\eta_c K$ with factorisation assumption from
CLEO~\cite{Edwards:2000bb}.  We now compare to other lattice QCD
calculations. The Jlab lattice group obtained~\cite{Dudek:2006ej}
$f_{J/\psi}$ = 399(4) MeV 
and $f_{\eta_c}$ = 429(4)(28) MeV from quenched
QCD, at a single lattice spacing. 
Chiu and Hsieh obtained $f_{\eta_c}$ = 438(5)(6) MeV
from a quenched QCD calculation using the overlap 
fermion action~\cite{Chiu:2007km},
at a single lattice spacing. 

\section{Conclusions}

We have described our lattice QCD calculations to compute the mass and
decay constants of the rho meson. We are trying to carefully study the
chiral extrapolation.  Similar to the chiral pertubation theory
studies with light pseudoscalar mesons, the errors on the masses and
decay constants have to be very precise to see the effects of loop
diagrams.

We reviewed why all three decay constants of the charmonium system are
important for 
the phenomenology of double charmonium production and we
presented preliminary results for these decay constants. We note that
the moments of the light cone wave functions could also be determined
in a similar manner to the calculation in the light quark
sector~\cite{Baron:2007ti}.

\acknowledgments

We thank Tassos Vladikas for discussions.


\end{document}